\documentclass[aps,pre,twocolumn,longbibliography,epsfig,floats,preprintnumbers,floatfix]{revtex4-2}
\usepackage{graphicx,amsmath,amssymb,bm,color,sidecap}
\usepackage[latin2]{inputenc}
\usepackage{dcolumn,nicefrac,blindtext,nicefrac,pgfplots}

\pgfplotsset{compat=1.18}
\usetikzlibrary{patterns}
\usetikzlibrary{3d}
\usetikzlibrary{math}
\usetikzlibrary{trees}
\usetikzlibrary{intersections, through, calc, }
\usetikzlibrary{graphs}
\usetikzlibrary{backgrounds,positioning,fit,petri, matrix}
\usetikzlibrary{arrows.meta}
\usetikzlibrary{decorations.footprints}
\usetikzlibrary{shapes.geometric}
\usetikzlibrary{shapes.symbols}
\usepgfplotslibrary{colormaps}
\usepgfplotslibrary{groupplots}
\usepgfplotslibrary{fillbetween}
\usepgfplotslibrary{external}
\tikzset{external/system call={lualatex --halt-on-error --jobname="\image" "\texsource" --}}
\tikzsetexternalprefix{Figs/tikz_external/}
\pgfplotsset{thesis_plot/.style={}}

\definecolor{farbe1}{HTML}{332288} 
\definecolor{farbe2}{HTML}{AA4499}  
\definecolor{farbe3}{HTML}{999933}  
\definecolor{farbe4}{HTML}{88CCEE}  

\definecolor{farbe4}{HTML}{fde725}
\definecolor{farbe3}{HTML}{35b779}
\definecolor{farbe2}{HTML}{31688e}
\definecolor{farbe1}{HTML}{440154}
\newcommand{\vek}[1]{\ensuremath\mathbf{#1}}

\newcommand{\abs}[1]{\ensuremath{\left|#1\right|}}
\newcommand{\norm}[1]{\ensuremath{\left\lVert#1\right\rVert}}
\newcommand{\sumnorm}[1]{\ensuremath{\left\lVert#1\right\rVert_{1}}}

\newcommand{\Erw}[1]{\ensuremath \left\langle{} #1 \right\rangle{}}

\newcommand{\integral}[5][]{%
  \ifthenelse{\equal{#1}{}}%
  {\ensuremath \int_{#2}^{#3}{ #4}\,\mathrm{d}{#5}}
  {\ensuremath \int_{#2}^{#3}{ #4}\,\mathrm{d}^{#1}{#5}}
}
\newcommand{\iintegral}[4][]{%
  \ifthenelse{\equal{#1}{}}%
  {\ensuremath \iint_{#2}{ #3}\,\mathrm{d}{#4}}
  {\ensuremath \iint_{#2}{ #3}\,\mathrm{d}^{#1}{#4}}
}
\newcommand{\iiintegral}[4][]{%
  \ifthenelse{\equal{#1}{}}%
  {\ensuremath \iiint_{#2}{ #3}\,\mathrm{d}{#4}}
  {\ensuremath \iiint_{#2}{ #3}\,\mathrm{d}^{#1}{#4}}
}

\newcommand{\ointegral}[5][]{%
  \ifthenelse{\equal{#1}{}}%
  {\ensuremath \oint_{#2}^{#3}{ #4}\,\mathrm{d}{#5}}
  {\ensuremath \oint_{#2}^{#3}{ #4}\,\mathrm{d}^{#1}{#5}}
}
\newcommand{\del}[0]{\ensuremath \partial}
\newcommand{\pdiff}[3][]{%
  \ifthenelse{\equal{#1}{}}%
  {\ensuremath \frac{\del #2}{\del #3}}
  {\ensuremath \frac{\del^{#1} #2}{\del #3^{#1}}}
}
\newcommand{\Pdiff}[3][]{
  \ifthenelse{\equal{#1}{}}%
  {\ensuremath \frac{\del}{\del #3} #2}
  {\ensuremath \frac{\del^{#1}}{\del #3^{#1}} #2}
}
\newcommand{\npdiff}[3][]{
  \ifthenelse{\equal{#1}{}}%
  {\ensuremath \nicefrac{\del #2}{\del #3}}
  {\ensuremath \nicefrac{\del^{#1} #2}{\del #3^{#1}}}
}
\newcommand{\pdel}[2][]{%
  \ifthenelse{\equal{#1}{}}%
  {\ensuremath \partial_{#2}}%
  {\ensuremath \partial_{#2}^{#1}}%
}

\newcommand{\diff}[3][]{%
  \ifthenelse{\equal{#1}{}}%
  {\ensuremath \frac{\mathrm d #2}{\mathrm d #3}}%
  {\ensuremath \frac{\mathrm d^{#1} #2}{\mathrm d {#3}^{#1}}}%
}

\newcommand{\iu}[0]{\ensuremath \mathfrak i}
\newcommand{\euler}[0]{\ensuremath \mathrm e}
\newcommand{\set}[2]{\ensuremath\left\{#1;\ #2\right\}}

\newcommand{\operator}[1]{\ensuremath\mathsf{#1}}

\newcommand{\ogfcorrespond}[0]{\ensuremath \overset{ \mathrm{ogf} }{\longleftrightarrow} }

\newcommand{\sequence}[2]{ \ensuremath \left( #1 \right)_{#2} }

\newcommand{\genmultinom}[2]{\ensuremath M_{#2}^{#1}}
\newcommand{\fallfak}[2]{#1^{\underline{#2}}}

\newcommand{\fakm}[2]{\Erw{\fallfak{#1}{#2}}}
\newcommand{\Li}[2]{\ensuremath\operatorname{Li}_{#1}\left(#2\right)}
\newcommand{\smallStirlingFirst}[2]{\ensuremath \left[\begin{smallmatrix} #1 \\ #2 \end{smallmatrix}\right]}
\newcommand{\smallStirlingSecond}[2]{\ensuremath \left\{\begin{smallmatrix} #1 \\ #2 \end{smallmatrix}\right\}}

\newcommand{\id}[0]{\operator{1}}
\pgfplotsset{compat=1.18}
\usetikzlibrary{patterns}
\usetikzlibrary{3d}
\usetikzlibrary{math}
\usetikzlibrary{trees}
\usetikzlibrary{intersections, through, graphs, calc, }
\usetikzlibrary{arrows,backgrounds,positioning,fit,petri, matrix}
\usetikzlibrary{decorations.pathmorphing, decorations.pathreplacing, decorations.shapes}
\usetikzlibrary{shapes.geometric}
\usetikzlibrary{shapes.multipart}
\usepgfplotslibrary{colormaps}
\usepgfplotslibrary{groupplots}
\usepgfplotslibrary{fillbetween}
\usepgfplotslibrary{external}
\tikzset{external/system call={lualatex --halt-on-error --output-format=dvi --jobname="\image" "\texsource" --}}
\tikzsetexternalprefix{Figs/tikz_external/}
\pgfplotsset{poster_plot/.style={
  poster_plot_scale,
  poster_plot_colorscheme,
}}
\pgfplotsset{poster_plot_scale/.style={
  small,
  x post scale = 0.75,
  y post scale = 0.75,
  font={\scriptsize},
  inner sep=0.3em,
}}
\pgfplotsset{poster_plot_colorscheme/.style={
  cycle list name = thesis_list,
  colorbar style={
    font={\scriptsize},
    inner sep=0.3em,
  },
  colormap name={inferno truncat},
}}
\pgfplotscreateplotcyclelist{thesis_list}{
  {farbe1, mark options={mark=*,         fill=farbe1!50, }, },
  {farbe2, mark options={mark=square*,   fill=farbe2!50, }, },
  {farbe3, mark options={mark=triangle*, fill=farbe3!50, }, },
  {farbe4, mark options={mark=diamond*,  fill=farbe4!50, }, },
  {farbe5, mark options={mark=pentagon*, fill=farbe5!50, }, },
  {farbe1, mark options={mark=otimes*,   fill=farbe1!50, }, },
  {farbe2, mark options={mark=oplus*,    fill=farbe2!50, }, },
  {farbe3, mark options={mark=star,      fill=farbe3!50, }, },
  {farbe4, mark options={mark=x,         fill=farbe4!50, }, },
  {farbe5, mark options={mark=+,         fill=farbe5!50, }, }% <-- don’t add a comma here
}
\pgfplotscreateplotcyclelist{thesis_list_hollow}{
  {farbe1, mark options={mark=o,        }, },
  {farbe2, mark options={mark=square,   }, },
  {farbe3, mark options={mark=triangle, }, },
  {farbe4, mark options={mark=diamond,  }, },
  {farbe5, mark options={mark=pentagon, }, },
  {farbe1, mark options={mark=otimes,   }, },
  {farbe2, mark options={mark=oplus,    }, },
  {farbe3, mark options={mark=star,     }, },
  {farbe4, mark options={mark=x,        }, },
  {farbe5, mark options={mark=+,        }, }% <-- don’t add a comma here
}
\pgfplotsset{frame-plot/.style={
  tiny,
  legend image code/.code={
    \draw [mark repeat=2,mark phase=2,##1]
      plot coordinates {
        (0cm,0cm)
        (0.15cm,0cm)
        (0.3cm,0cm)
      };
  },
  very axis title/.append style={font={\tiny}},
  colormap={thermal2}{rgb255=(0,0,0) rgb255=(77,0,179) rgb255=(255,51,0) rgb255=(255,255,0) },
  cycle list name = thesis_list,
  colorbar style={
    x post scale=0.5,
    every axis title/.append style={font={\tiny}},
    font={\tiny},
  },
}}
\pgfplotsset{thesis_plot_scale/.style={
  x post scale=0.75,
  y post scale=0.75,
  legend cell align=left,
  legend style={legend style={font=\scriptsize}},
}}
\pgfplotsset{thesis_plot_colorscheme/.style={
  colormap={thermal2}{rgb255=(0,0,0) rgb255=(77,0,179) rgb255=(255,51,0) rgb255=(255,255,0) },
  colormap name={inferno truncat},
  cycle list name = thesis_list,
}}
\pgfplotsset{thesis_plot_normalsize/.style={
  thesis_plot_colorscheme,
}}
\pgfplotsset{thesis_plot/.style={
  thesis_plot_normalsize,
  thesis_plot_scale,
}}
\pgfplotsset{paper_ticks/.style={
  tick style={
    thin,
  },
}}
\pgfplotsset{paper_plot/.style={
  thesis_plot,
  paper_ticks,
}}
\pgfplotsset{colorbar top/.style={
  colorbar horizontal,
  colorbar style={
    at={(1,1.03)},
    anchor=south east,
    xticklabel pos=upper,
  },
}}
\pgfplotsset{colorbar shorter/.style={
  colorbar style={
    width=0.6*\pgfkeysvalueof{/pgfplots/parent axis width},
  },
}}
\pgfplotsset{
  every axis legend/.append style={
    outer sep=3pt,
  },
}

\begin{document}
\title{Inferring Tree Structure with Hidden Traps from First Passage Times}
\author{Fabian H.\ Kreten}
\affiliation{Department of Theoretical Physics and Center for Biophysics, 
Saarland University, 66123 Saarbr\"ucken, Germany}
\author{Ludger Santen}
\affiliation{Department of Theoretical Physics and Center for Biophysics, 
Saarland University, 66123 Saarbr\"ucken, Germany}
\author{Reza Shaebani}
\affiliation{Department of Theoretical Physics and Center for Biophysics, 
Saarland University, 66123 Saarbr\"ucken, Germany}

\begin{abstract}
Tracking the movement of tracer particles has long been a strategy for uncovering complex 
structures. Here, we study discrete-time random walks on finite Cayley trees to infer key 
parameters such as tree depth and geometric bias toward the root or leaves. By analyzing 
first passage properties, we show that the first two first-passage-time factorial moments 
(FPTFMs) uniquely determine the tree structure. However, if the random walker experiences 
waiting phases---due to sticky branch walls or presence of traps---this identification 
becomes nontrivial. We demonstrate that the generating function of the first passage time 
(FPT) distribution decomposes into contributions from the waiting time distribution and 
the random walk without waiting, leading to a nonlinear system of equations relating the 
factorial moments of the waiting time distribution and the FPTFMs of random walks with 
and without waiting. For geometrically distributed waiting times, additional moment 
measurements do not suffice, but unique determination of the structure is achieved 
by varying initial conditions or fitting the Fourier transform of the FPT distribution 
to measured data. The latter method remains effective also for power-law waiting time 
distributions, where higher-order FPTFMs are undefined. These results provide a framework 
for reconstructing tree-like networks from FPT data, with applications in biological 
transport and spatial networks.
\end{abstract}

\maketitle

\section{Introduction}
\label{Sec:Introduction}

Tracking the movement of tracer particles has long been a practical method for probing 
the unknown structure and topology of labyrinthine environments, ranging from disordered 
media to biological transport networks \cite{benAvraham00,Krishna09,Mair99,Mitra92,Chen12,
Tierno16,Kac66,Felici04,Sadjadi08,Durian91,Maret97,Cooper16,Agliari07,Jose18}. For 
instance, diffusion propagators have been used to estimate structural characteristics of 
porous media, including porosity, confinement, permeability, absorption strength, and 
surface-to-volume ratio \cite{Chen12,Krishna09,Mitra92,Mair99}. Other examples include 
identifying magnetic bubble arrangements through the anomalous behavior of the mean square 
displacement of paramagnetic colloids in flashing magnetic potentials \cite{Tierno16}, or 
determining the geometry of absorbing boundaries---such as those in acoustic cavities---from 
the eigenvalue spectrum of the confined diffusion equation \cite{Kac66}.

Analyzing diffusive dynamics, e.g.\ by measuring the diffusion constant or 
mean square displacement, typically requires direct tracking of the tracer, which 
can pose technical challenges or necessitate invasive methods in biological or medical 
contexts. Moreover, such measures often fail to capture fine structural details. As 
an alternative, other transport quantities have been employed to indirectly assess 
the structural characteristics of interest. For instance, the absorption efficiency 
of diffusing oxygen can reflect the topology of the bronchial tree in the human lung 
\cite{Felici04}, or the mean free path of light relates to obstacle size and density 
in turbid media \cite{Sadjadi08}, enabling the use of diffusive light propagation to 
probe the temporal structural evolution of foams and opaque environments \cite{Durian91,
Maret97}. Among other measurable quantities, first passage time (FPT) properties offer 
promising tools for indirect structural analysis. For example, the first return time 
of random walks has been used to estimate geometrical properties of complex networks, 
including the number of triangles, loops, and subgraphs \cite{Cooper16}. Similarly, 
the mean time for reactants to reach a reaction center or to encounter each other 
can predict the timescale of autocatalytic reactions on inhomogeneous substrates 
\cite{Agliari07}.

\begin{figure*}
\centering
\includegraphics[width=0.8\textwidth]{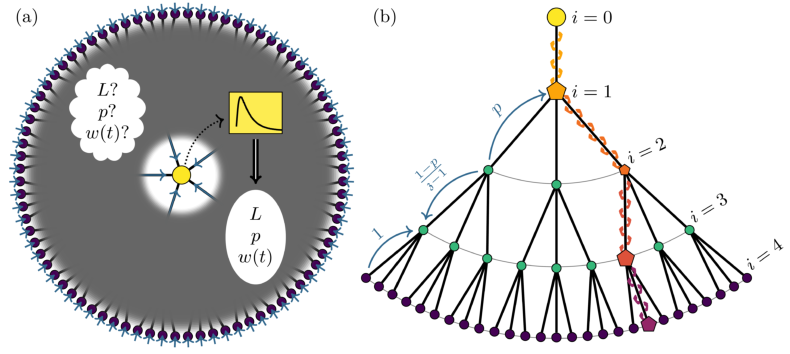}
\caption{(color online) Schematic illustration of the problem and modeling framework. 
(a) A Cayley tree of unknown structure (represented by the gray shading), characterized 
by its depth $L$, upward hopping probability $p$ (in conjunction with the coordination 
number $\mathfrak z$), and the distribution $w(t)$ of waiting times induced by stickiness 
or hidden traps along the branch walls. Random walkers are released from a specific 
level of the tree---here, the leaf nodes (dark purple)---and eventually reach the root 
(yellow). The distribution of their first passage times (inset graph) is used to infer 
the structural parameters $L$, $p$ and $w(t)$. (b) A single subtree of a Cayley tree of 
depth $L=4$ and coordination number $\mathfrak z=4$. The leaf nodes ($i=L=4$) are 
displayed in dark purple, the root node ($i=0$) in yellow. Transition probabilities 
are indicated on the left. On the right, a sample trajectory of a walker moving from 
a leaf to the root is depicted with footsteps and pentagons; the size of each pentagon 
reflects the random waiting time at that node, while the color of the pentagons and 
footsteps transitions from purple to yellow to indicate the progression of time. A 
cycle-free sample path is chosen to clearly illustrate the waiting behavior.}
\label{Fig:1}
\end{figure*}

Our study focuses on tree-like structures, which represent a broad range of real-world 
systems---from synthetic polymer configurations and dendrimer macromolecules to the 
bronchial architecture of lungs, vascular networks, neuronal dendrites, as well as 
certain communication and power distribution networks, and river basins \cite{Felici04,
Spruston08,Hering01,Jose18,Vetter01,Kreten24,Rodriguez01,Banavar99,Helfand83,Wu12,
Heijs04,Katsoulis02,Argyrakis00,Fleury01}. Tree-like architectures are also highly 
relevant in network and graph theory, epidemic modeling, and computational search 
algorithms \cite{Newman10,PastorSatorras15,Kleinberg00,Szabo02,Bollobas04}. Regular 
finite Cayley trees and infinite Bethe lattices have been thoroughly studied by mapping 
them onto effective 1D lattices \cite{Agliari08,Metzler14,Shaebani18,Redner01,Hughes82,
Monthus96,Cassi89,Khantha83,Skarpalezos13}, enabling the calculation of stochastic 
quantities of interest, such as the probability of return to the origin and the mean 
first passage time (MFPT) to reach a target node. The latter has been shown to depend 
on both the size of the tree and the bias in hopping toward the target. Consequently, 
measuring the MFPT alone is typically insufficient to uniquely determine the structure 
of even simple Cayley trees. While it has been proposed that using random walkers with 
varying waiting time statistics may aid in structural inference \cite{Shaebani18}, such 
waiting is often an inherent property of the system---originating from stickiness or 
hidden traps along the tree branches---rather than a characteristic of the random 
walker itself. Here, we are particularly interested in predicting the structure of 
more intricate, nonuniform (depth-dependent) trees---such as those that thin or 
thicken toward the root---and under the influence of inherent stochastic waiting 
events along the branches. We adopt discrete-time random walks on finite Cayley 
trees as an analytically tractable framework for studying such branched structures. 
While continuous-time random walks are a widely used alternative, we focus on 
a discrete-time approach, which naturally captures stepwise processes, such as 
sequential movement in biological systems or signal propagation in networks.

We employ a FPT-based analysis as a powerful tool for inferring structural parameters 
of the system---such as the bias in movement toward the root or the frequency of 
temporary sticking and trapping along the path---as illustrated in Fig.\,\ref{Fig:1}(a). 
We demonstrate that knowing two first-passage-time factorial moments (FPTFMs) for 
tracer particles to reach the root of a Cayley tree is sufficient to uniquely 
determine the structure of the tree in the absence of traps. We then extend our 
approach to incorporate waiting times resulting from the presence of traps. In 
this case, the generating function of the FPT distribution of the random walk can 
be written as the composition of the generating function of the FPT distribution 
of a walk without delay and the generating function of the waiting time distribution. 
This decomposition can yield a nonlinear system of equations that relates factorial 
moments of the waiting time distribution and the FPTFMs of walks with and without 
trapping effects. We show that accurate inference of the tree structure in such 
cases requires more refined analysis---such as probing with varied initial conditions 
or fitting the Fourier transform of the FPT distribution to observed data. Notably, 
our method remains effective even in the presence of power-law distributed waiting 
times, where higher-order FPTFMs may diverge.

The remainder of the paper is organized as follows. In Section~\ref{Sec:Method}, 
we introduce the mathematical formulation of the problem and derive the generating 
function relations for FPTs without and with waiting times, respectively. 
Sections~\ref{Sec:RW_without_waiting} and \ref{Sec:RW_with_waiting} consider 
the FPTFM-based inference of structural parameters in absence and presence 
of waiting. Section \ref{Sec:Fourier_method} presents the inference framework 
using the Fourier transform of the FPT distribution and examines the impact 
of different waiting time distributions. Finally, Sec.~\ref{Sec:Conclusion} 
summarizes the main findings and concludes the paper.

\section{Method}
\label{Sec:Method}

In this section, we establish the mathematical framework underlying our analysis. 
A central tool throughout is the use of generating functions, widely employed in 
the study of combinatorial sequences \cite{Riordan80,Wilf94,Flajolet09} and the 
probability distributions of discrete random variables \cite{VanKampen11}. Given 
a sequence $\sequence{a(n)}{n\geq 0}$, its ordinary generating function is 
defined by the power series $\hat a(z) = \sum_{n=0}^\infty a(n)\,z^n$. Throughout 
this work, we use a hat to denote the generating function of the corresponding 
quantity. We use the symbol $\ogfcorrespond$ to indicate the correspondence 
between relations in the sequence domain and their ordinary generating function 
counterparts. A particularly useful property of ordinary generating functions 
is the convolution property \cite{Wilf94}, which states that for $k$ sequences
$\sequence{a_1(n)}{n\geq 0},\ \ldots,\ \sequence{a_k(n)}{n\geq 0}$, the 
convolution satisfies
\begin{equation}
b(n)=\!\!\sum_{n_1+\ldots+n_k=n}\!\! a_1(n_1)\cdots a_k(n_k) \ogfcorrespond{} 
\hat b(z) = \prod_{i=1}^k \hat a_i(z),
\label{Eq:Convolution}
\end{equation}
where the sum runs over all $k$-tuples $(n_1,\ \ldots,\ n_k) \in \mathbf N^k$ 
such that $n_1\,{+}\,\cdots\,{+}\,n_k\,{=}\,n$.

The second key tool used in this paper is the factorial moment, a well-established 
quantity for studying nonnegative integer-valued random variables and their distributions 
\cite{Riordan80,VanKampen11}. Factorial moments have also seen applications in data 
analysis, particularly in high-energy physics \cite{Bialas86}. Let $\fallfak{t}{k} 
= \prod_{n=0}^{k-1} (t-n)$ denote the $k$th falling factorial, with the convention 
$\fallfak t 0 = 1$. Then, for a random variable $t$ with the probability distribution 
$f(t)$ the expectation value of the $k$th falling factorial defines the $k$th factorial 
moment $\fakm{t}{k}$ of $t$. A property which makes factorial moments more convenient 
that ordinary moments is that they are directly generated by derivatives of the generating 
function $\hat f(z)$ of the probability distribution:
\begin{equation}
\fakm{t}{k} = \left.\diff[k]{\hat f}{z}\right|_{z=1}.
\label{Eq:FM}
\end{equation}
Since monomials can be expressed in terms of falling factorials, factorial and ordinary 
moments, $\fakm t n$ and $\Erw{t^n}$, are related to each other via $\fakm t n {=} 
\sum_{k=0}^n (-1)^{n-k}\smallStirlingFirst n k \Erw{t^k}$ and $\Erw{t^n} {=} 
\sum_{k=0}^n \smallStirlingSecond n k \fakm t k$, where $\smallStirlingFirst n k$ 
and $\smallStirlingSecond m k$ denote the (unsigned) Stirling numbers of first and 
second kind \cite{Flajolet09}.

\subsection{Recurrence relation for FPTFMs}

Having introduced factorial moments, we now derive a recurrence relation for the 
FPTFMs. Notably, this derivation does not rely on any assumptions about the random 
walk taking place on a tree structure; it holds for arbitrary discrete random 
walks on possibly infinite state spaces. Furthermore, extending the approach to 
continuous spaces is straightforward and likely requires only minimal adaptation.

A well-established approach for obtaining the first passage time distribution 
$f(t)$ is through the survival probability $S(t)$, which denotes the probability 
that the walker has not reached the target by time $t$. The relation between 
these two quantities, as well as between their generating functions, is given by
$f(t+1) = S(t) -S(t+1) \ogfcorrespond{} \hat f(z) = (z -1)\hat S(z) +1$ \cite{Redner01}.
Applying Eq.\,(\ref{Eq:FM}) to this relation allows one to express the FPTFMs 
in terms of the survival probability:
\begin{equation}
\frac 1 {k}\fakm{t}{k} = \left.\diff[k-1]{}{z}\hat S\right|_{z=1}.
\label{Eq:FM_S}
\end{equation}

This framework can be directly applied to Markovian random walks in discrete time, 
as considered in the following. Let $\vek P(t)$ denote the vector (or sequence) of 
occupation probabilities for all states at time $t$. The time evolution of $\vek P$ 
is governed by the Chapman-Kolmogorov equation $\vek P(t+1) = \operator M\,\vek P(t)$, 
where $\operator M$ is the transition matrix. If $\operator M$ respects the absorbing 
nature of the target sites, then the survival probability vector $\vek S(t)$, whose 
entries correspond to the survival probability from each starting site, evolves 
according to the backward equation $\vek S(t+1) = \operator M^\dagger \vek S(t) 
\ogfcorrespond{} -1 = (z\operator M^\dagger -\operator 1) \hat{\vek S}(z)$, where 
$\operator M^\dagger$ is the transpose of $\operator M$. Applying Eq.\,(\ref{Eq:FM_S}) 
to this relation yields the following recurrence for the FPTFMs:
\begin{equation}
-k \operator M^\dagger \fakm{\vek t}{k -1} = (\operator M^\dagger -\id)\fakm{\vek t}{k} 
\text{ for } k > 0,
\label{Eq:FM_recurrence} 
\end{equation}
where $\fakm{\vek t}{k -1}$ is the vector of FPTFMs indexed over the starting sites. 
This relation serves as the discrete-time analog of a well-known identity for ordinary 
moments in continuous-time processes \cite{Risken96}.

\subsection{Decoupling the FPT distribution for walks with waiting}

Next, we derive the relation between the ordinary generating function of the FPT 
distribution for a discrete-time random walk with site-independent waiting and 
that of a walk without waiting, along with the generating function of the waiting 
time distribution. This decoupling approach parallels well-known results obtained 
in continuous time and space using Laplace transforms \cite{Dienst07}, and in discrete space 
using cumulant generating functions \cite{Manhart15}, but is here expressed directly 
in terms of ordinary generating functions.

Let $f(t)$ denote the FPT distribution for a random walk with independent and identically 
distributed (i.i.d.) waiting times drawn from the distribution $w(t)$ at each site. Let 
$\mathfrak{f}(t)$ be the FPT distribution of the same walk without waiting times (i.e., 
$w(t) \,{=}\, \delta_{t,1}$). Then, $f(t)$ is given by
\begin{equation}
f(t) = \sum_{k = 0}^\infty \mathfrak{f}(k) \sum_{\tau_1 +\dots +\tau_k = t} w(\tau_1) 
\cdots w(\tau_k), 
\end{equation}
and applying the convolution property of ordinary generating functions, 
Eq.\,(\ref{Eq:Convolution}), the generating function of $f(t)$ becomes:
\begin{equation}
\hat f(z) = \sum_{k=0}^\infty \mathfrak{f}(k) \hat w(z)^k = \hat{\mathfrak f}(\hat w(z)),
\label{Eq:GF_FPT_Waiting}
\end{equation}
i.e., the composition of the ordinary generating functions of the FPT distribution
without waiting and the waiting time distribution.

Applying Eq.\,(\ref{Eq:FM_S}) to Eq.\,(\ref{Eq:GF_FPT_Waiting}) and using the normalization 
condition $\hat w(1) \,{=}\, 1$ for probability distributions, one can express the 
FPTFMs with waiting $\fakm{t}{k}$ in terms of the FPTFMs without waiting $\fakm{
\mathfrak{t}}{k}$ and the factorial moments of the waiting time distribution 
$\fakm{\tau_w}{k}$:
\begin{equation}
\begin{aligned}
\fakm{t}{1} &= \fakm{\mathfrak{t}}{1}\fakm{\tau_w}{1}, \\
\fakm{t}{2} &= \fakm{\mathfrak{t}}{2}\fakm{\tau_w}{1}^2 
+\fakm{\mathfrak{t}}{1}\fakm{\tau_w}{2}, \\
\fakm{t}{3} &= \fakm{\mathfrak{t}}{3}\fakm{\tau_w}{1}^3 
+3\fakm{\mathfrak{t}}{2}\fakm{\tau_w}{1}\fakm{\tau_w}{2} 
+\fakm{\mathfrak{t}}{1}\fakm{\tau_w}{3}, \\
    &\vdots \\
\fakm{t}{m} &= \sum_{\vek b \in B_m} \genmultinom{m}{\vek b} 
\fakm{\mathfrak{t}}{\sumnorm{\vek b}} \prod_{k=1}^{m}
\left(\frac{\fakm{\tau_w}{k}}{k!}\right)^{b_k}.
\end{aligned}
\label{Eq:FM_Waiting_first3}
\end{equation}
The last line represents the general form obtained using Fa\`a di Bruno's formula; 
see Appendix for details. Here $B_m = \set{\vek b \in \mathbf N^m}{ \sum_{k=1}^m k b_k 
= m}$, $\genmultinom{m}{\vek b} = \frac{m!}{\prod_{k=1}^m b_k!}$ a generalized 
multinomial coefficient \footnote{The multinomial coefficient $\binom{m}{\vek b}$ 
has the same formula as $\genmultinom{m}{\vek b}$, but also requires $m = 
\sumnorm{\vek b}$.}, and $\sumnorm{\vek b} = \sum_{k=1}^m b_k$. We note that 
$\sumnorm{\vek b} \leq m$, with equality $\sumnorm{\vek b} = m$ if and only if 
$\vek b = (m, 0, \ldots, 0)$. Due to its complexity, the general expression is 
primarily useful for theoretical purposes, while the finite number of moments 
typically required in applications can be computed manually or with the aid 
of a computer algebra system. In the following, we use the unified notation 
$\fakm{\tau_w}{k}$ for the factorial moments of the waiting time distribution, 
$\fakm{t}{k}$ for the FPTFMs with waiting, and $\fakm{\mathfrak{t}}{k}$ for 
the FPTFMs without waiting. Moreover, the index $i$ is used for the FPTFMs 
with/without waiting, $\fakm{t_i}{k}$ and $\fakm{\mathfrak t_i}{k}$, if 
starting from level $i$.

\begin{figure*}
\centering
\includegraphics[width=0.99\textwidth]{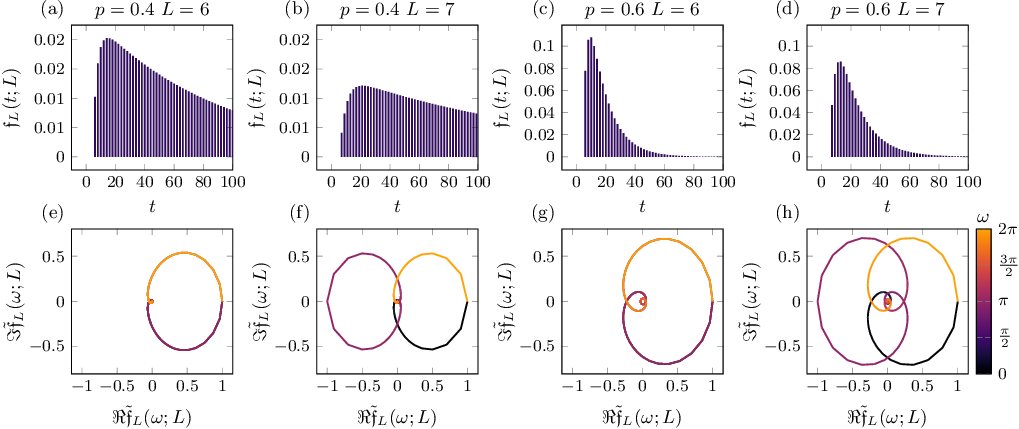}
\caption{(color online) First passage time distributions and their Fourier transforms for random 
walks on regular trees without waiting. (a-d) Examples of the FPT distribution 
$\mathfrak f_L(t;L)$ from the leaves to the root of a regular Cayley tree, for a random walker 
without waiting times. $\mathfrak f_L(t;L)$ is nonzero for $t\,{\geq}\,L$ and only for even or 
odd $t$ depending on $L$. (e-h) Fourier transforms $\tilde{\mathfrak f}_L(\omega; L)$ of the 
corresponding FPT distributions (from panels a-d), shown in the complex plane. 
Each contour encloses the origin and loops exactly $L$ times. For even values 
of $L$, the contour is traced twice. The parameters $L$ and $p$ used for each 
column are indicated at the top.}
\label{Fig:2}
\end{figure*}

\subsection{Cayley trees and random walks thereon}

To conclude this section, we introduce Cayley trees and apply the general results 
of the preceding subsections to random walks on such structures. We then outline 
the waiting time distributions considered in this work and comment on the connection 
between the generating function and the Fourier transform of the FPT distribution.

The random walk takes place on a finite Cayley tree of depth $L$, which is a 
regular cycle-free graph constructed as follows: The root node is assigned to 
level or shell $i\,{=}\,0$. It connects to $\mathfrak z$ child nodes, forming 
level $i\,{=}\,1$. For levels $0\,{<}\,i\,{<}\,L$, each node at level $i$ is 
connected to $\mathfrak z\,{-}\,1$ nodes forming the level $i\,{+}\,1$. The 
parameter $\mathfrak z$ denotes the coordination number of the tree \cite{Ostilli12}; 
see Fig.\,\ref{Fig:1}(b).

The random walk proceeds as follows: At each node on levels $0\,{<}\,i\,{<}\,L$, 
the walker moves to the parent node at level $i\,{-}\,1$ with probability $p$, 
and to each of the $\mathfrak z\,{-}\,1$ child nodes at level $i\,{+}\,1$ with 
probability $\frac{1-p}{\mathfrak z -1}$. At the leaves ($i\,{=}\,L$), the walker 
always moves to the parent level ($i\,{=}\,L\,{-}\,1$) with probability $1$. The 
dynamics at the root ($i\,{=}\,0$) are irrelevant for computing FPTs to it, and 
we treat it as an absorbing state to facilitate the use of the adjoint (backward) 
equation formalism. For unbiased diffusion on a Cayley tree, the rootward transition 
probability $p$ is simply related to the branching parameter $\mathfrak z$ by 
$p\,{=}\,\frac{1}{\mathfrak z}$. In the presence of a bias or an energy difference 
$\Delta E$ between consecutive levels, this relation is modified to $p\,{=}\,\frac{1}{
1\,{+}\,(\mathfrak z\,{-}\,1)\exp(-\Delta E{/}k_B\,T)}$, with $k_B$ denoting 
Boltzmann constant \cite{Heijs04}. Possible sources of such bias include shortest-path 
preferences or morphological tapering, as observed in dendritic trees of neurons 
\cite{Jose18}. For this reason, we treat $p$ as a tunable parameter, which allows 
the formalism to remain general and applicable to both unbiased and biased scenarios.

Since our primary interest lies in the first passage time to the root, and the 
transition probabilities are identical for all nodes at a given level, the dynamics 
can be effectively reduced to a one-dimensional process over levels rather than 
individual nodes \cite{Agliari08,Metzler14,Shaebani18,Redner01,Hughes82,Monthus96,
Cassi89,Khantha83,Skarpalezos13}. This reduction renders the transition matrix
\begin{equation}
\operator M =
\begin{pmatrix}
    1 &   p &        &        &     &   \\
    0 &   0 &      p &        &     &   \\
      & 1-p &      0 & \ddots &     &   \\
      &     & \ddots & \ddots &   p &   \\
      &     &        &    1-p &   0 & 1 \\
      &     &        &        & 1-p & 0 \\
\end{pmatrix}
\label{Eq:Transport_Operator}
\end{equation}
tridiagonal, significantly simplifying the computation of the FPTFMs $\fakm{\mathfrak{t_i}}{k}$ 
from any level $i$ to the root in $\mathcal O(L\,k)$ time using Eq.\,(\ref{Eq:FM_recurrence}). 
For target sites other than the root, reduction to one dimension is not possible 
anymore and the transition matrix for the full tree must be considered.

Furthermore, the generating function of the FPT probability distribution 
$\hat{\mathfrak f}_L(z; L) = \sum_{t=0}^\infty \mathfrak f_L(t; L) z^t$ 
to the root for a random walker starting at level $L$ (the leaves, denoted by the index) of 
a tree of depth $L$ (denoted by the second argument) has been derived previously 
\cite{Jose18,Shaebani18}. Adapting their result to our setting without waiting 
times, and using the abbreviation $A(z)=\sqrt{1-4 (1-p) p z^2}$, the expression 
reads
\begin{widetext}
\begin{equation}
\hat{\mathfrak f}_L (z; L) = \frac{2^{L+1} p^{L} z^{L} A(z)}{(1 +A(z))^{L} (2p -(1 -A(z))) 
-(1 -A(z))^{L} (2p -(1 +A(z)))}.
\label{Eq:GF_Leaves}
\end{equation}
\end{widetext}
Since $A(z)$ is an even function and the denominator of $\hat{\mathfrak f}_L(z;L)$ is odd in $A$, 
$\hat{\mathfrak f}(z;L)$ has the same parity as $L$. Consequently, the FPT probability distribution 
$\mathfrak f_L(t; L)$ vanishes for all $t$ that are not of the same parity as $L$. This behavior 
is expected because the tree with the transition rules defined in 
Eq.\,(\ref{Eq:Transport_Operator}) forms a bipartite network and the walker alternates 
between the two parts in every step. Hence, it requires an even number of steps 
to reach the root once it has entered the part containing the root.

Note that for $L\,{>}\,1$, the decomposition of the time to reach the root starting 
from the leaves into the time to reach the level below the root starting from the 
leaves and the time to reach the root starting from the level below, i.e.\ 
$T_{L\rightarrow 0} = T_{L\rightarrow 1} +T_{1\rightarrow 0}$, holds and that 
$T_{L\rightarrow 1}$ and $T_{1\rightarrow 0}$ are distributed according to 
$\mathfrak f_{L-1}(t; L-1)$ and $\mathfrak f_1(t; L)$. Using Eqs.\,(\ref{Eq:Convolution}) and 
(\ref{Eq:GF_Leaves}), the generating function of the FPT distribution to reach 
the root starting from the level below can be obtained as
\begin{equation}
\hat{\mathfrak f}_1(z; L) = \frac{\hat{\mathfrak f}_L(z; L)}{\hat{\mathfrak f}_{L-1}(z; L -1)}.
\label{Eq:GF_Level1}
\end{equation}
Using Eqs.\,(\ref{Eq:FM}), (\ref{Eq:GF_Leaves}), and (\ref{Eq:GF_Level1}), one can 
in principle derive explicit expressions for the corresponding factorial moments. 
However, due to the complexity of the resulting formulas, we do not pursue this 
route further here.

We consider two waiting time distributions in this paper, both supported on the 
set of positive integers $t \,{\geq}\, 1$. The first one is the geometric distribution
\begin{equation}
w(t) = q(1-q)^{t -1} \ogfcorrespond{} \hat w(z) = \frac{q z}{1 -(1-q)z}
\label{Eq:Geometric_Waiting}
\end{equation}
which arises in the context of random walks with spontaneous stepping with probability 
$q$. The second one is the zeta distribution, defined by
\begin{equation}
w(t) = \frac{t^{-s}}{\Li{s}{1}} \ogfcorrespond{} \hat w(z) = \frac{\Li{s}{z}}{\Li{s}{1}},
\label{Eq:Zeta_Waiting}
\end{equation}
where $s \,{>}\, 1$ and $\Li s z = \sum_{t\geq 1} t^{-s} z^t$ denotes the Polylogarithm 
function. This distribution is considered in Sec.\,\ref{Sec:Fourier_method}. For the 
zeta distribution, only the (factorial) moments up to order $s\,{-}\,1$ are finite. 
In the numerical calculations presented in Sec.\,\ref{Sec:Fourier_method}, the 
polylogarithm function must be evaluated for real parameters $s$ and complex 
arguments $z$; suitable numerical libraries are available for this purpose 
\cite{Roughan20}. 

For the analysis in Sec.\,\ref{Sec:Fourier_method}, the discrete-time Fourier 
transform (DTFT) of the FPT distribution $\tilde f(\omega)$ is required. This 
quantity is readily obtained from the generating function $\hat f(z)$ of the FPT 
distribution via $\tilde f(\omega) = \sum_{t\geq 0} f(t) \euler^{-\iu\omega t} = 
\hat f\left(\euler^{-\iu\omega}\right)$. Moreover, sampling the DTFT at regularly 
spaced points $k\,{=}\,0, \ldots, N\,{-}\,1$ around the unit circle and applying 
the inverse discrete time Fourier transform provides a method to numerically 
invert the generating function and recover the FPT distribution \cite{Mills87,
Horvath20}. Figure\,\ref{Fig:2} shows representative FPT distributions and their 
corresponding DTFTs for various parameter sets.

Each contour in panels (e)-(h) of Fig.\,\ref{Fig:2} encloses the origin and loops 
exactly $L$ times, reflecting the depth of the tree. This can be understood by 
a continuity argument. For a tree of depth $L$ with walkers starting from the 
leaves, the earliest possible arrival time is $t\,{=}\,L$ and the generating 
function of the FPT distribution $\hat{\mathfrak f}_L(z;L)$ has its lowest nonzero 
contribution at order $z^L$. Hence, for small radii $\rho\,{\ll}\,1$ the image 
of the circle $|z|\,{=}\,\rho$ under $\hat{\mathfrak f}_L(z;L)$ is approximately 
parameterized by $s \,{\mapsto}\, \text{e}^{-\text{i} L s}$, which winds $L$ 
times around the origin as $s$ varies from $0$ to $2\pi$. By continuity, 
increasing $\rho$ up to $1$ leaves this winding number unchanged, which 
explains why the contours---corresponding to images of the unit circle under 
$\hat{\mathfrak f}_L(z;L)$---in Fig.\,2(e)-(h) loop exactly $L$ times.

\begin{figure}[t]
\centering
\includegraphics[width=0.45\textwidth]{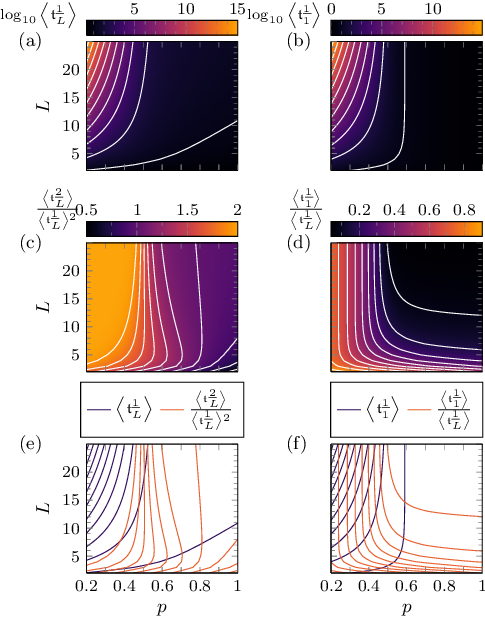}
\caption{(color online) First-passage-time factorial moments as functions of tree depth $L$ and 
upward hopping bias $p$. (a,b) Logarithm of the first factorial moment for walkers 
starting from the leaf nodes ($\log_{10}\fakm{\mathfrak{t}_L}{1}$, panel a) or 
starting one level below the root ($\log_{10}\fakm{\mathfrak{t}_1}{1}$, panel b). 
Lines are contours of constant $\log_{10}\fakm{\mathfrak{t}_L}{1}$ or 
$\log_{10}\fakm{\mathfrak{t}_1}{1}$ respectively. (c) Normalized second factorial 
moment $\frac{\fakm{\mathfrak{t}_L}{2}}{\fakm{\mathfrak{t}_L}{1}^2}$ for walkers 
starting from the leaves. This quantity saturates at large $L$ and small $p$, 
limiting its utility for parameter inference in that regime. (d) Ratio 
$\frac{\fakm{\mathfrak{t}_1}{L}}{\fakm{\mathfrak{t}_L}{1}}$, comparing first 
factorial moments for walkers starting just below the root versus from the leaves. 
This ratio becomes independent of $L$ for small $p$ and sufficiently deep trees, 
making it a robust alternative for inferring tree parameters when the normalized 
second moment is saturated. (e,f) Contour plots showing combinations of moment-based 
observables. In (e), contours of $\log_{10} \fakm{\mathfrak{t}_L}{1}$ and 
$\frac{\fakm{\mathfrak{t}_L}{2}}{\fakm{\mathfrak{t}_L}{1}^2}$ intersect at a 
unique point corresponding to the true values of $L$ and $p$. Similarly, (f) 
shows that intersections of $\log_{10}\fakm{\mathfrak{t}_1}{1}$ and 
$\frac{\fakm{\mathfrak{t}_1}{L}}{\fakm{\mathfrak{t}_L}{1}}$ contours provide 
an alternative route to identify the tree parameters.}
\label{Fig:3}
\end{figure}

\section{Random walk without waiting}
\label{Sec:RW_without_waiting}

With the theoretical groundwork established, we now turn to the task of inferring 
structural properties of Cayley trees from FPT statistics. We begin with the 
case of a random walk without waiting times. In this setting, the phase space 
of the Cayley tree is fully determined by two parameters: the tree depth $L$, 
and the bias probability $p$, which may implicitly encode the coordination 
number $\mathfrak z$. Consequently, the FPT distribution and its moments 
depend only on $L$ and $p$. This implies that, in principle, two independent 
observable derived from FPT statistics should suffice to uniquely identify 
these parameters.

In Fig.\,\ref{Fig:3}(a), the phase diagram of the first factorial moment $\fakm{\mathfrak t_L}{1}$ 
is shown, and in Fig.\,\ref{Fig:3}(c), the normalized second factorial moment 
$\frac{\fakm{\mathfrak t_L} 2}{\fakm{\mathfrak t_L} 1^2}$ is plotted for random walkers starting 
from the leaves. Across most of the phase space, the contour lines of constant 
$\fakm{\mathfrak t_L} 1$ and $\frac{\fakm{\mathfrak t_L} 2}{\fakm{\mathfrak t_L} 1^2}$ intersect at a unique 
point, which identifies the corresponding parameters $p$ and $L$; c.f.\ 
Fig.\,\ref{Fig:3}(e). The only exception occurs in the regime of low $p$ 
and high $L$ where $\frac{\fakm{\mathfrak t_L} 2}{\fakm{\mathfrak t_L} 1^2}$ saturates. 
This ambiguity can be resolved by additionally considering the first FPTFM 
$\fakm{\mathfrak t_1} 1$. Including another initial condition introduces additional 
overhead in measuring the factorial moments, as it requires conducting an 
extra experiment for each added initial condition. The relation $\fakm{\mathfrak t_{L,L}} 1 
= \fakm{\mathfrak t_{1,L}} 1 +\fakm{\mathfrak t_{L-1,L-1}} 1$ (which follows from Eq.\,(\ref{Eq:GF_Level1})) 
links the first FPTFM from the level below the root to those from the leaves 
in trees of depths $L-1$ and $L$ as denoted by the second index. Moreover, if particles are allowed to leave 
the root, $\fakm{\mathfrak t_1}{1}$ corresponds to the mean return time to the root minus 
the mean time it takes to leave it. Together, these relations offer a way to 
avoid conducting a second experiment, provided it is possible to determine 
whether a particle has already reached the root or the level below. This 
could be achieved, for instance, by marking or using distinguishable 
particles.

Figures \ref{Fig:3}(b) and (d) show phase diagrams of the first FPTFM 
$\fakm{\mathfrak t_1} 1$ (starting from the level below the root) and the ratio 
$\frac{\fakm{\mathfrak t_1}{1}}{\fakm{\mathfrak t_L}{1}}$. When plotting the contour lines 
of $\fakm{\mathfrak t_1}{1}$ and $\frac{\fakm{\mathfrak t_1}{1}}{\fakm{\mathfrak t_L}{1}}$ in a single 
coordinate system, as in Fig.\,\ref{Fig:3}(f), they intersect at unique 
points---even in the regime where $\frac{\fakm{\mathfrak t_L}{2}}{\fakm{\mathfrak t_L}{1}^2}$
saturates. Since $\fakm{\mathfrak t_1}1$ and $\fakm{\mathfrak t_L} 1$ exhibit similar dependence 
on the parameters, they can be used interchangeably in the preceding inference 
framework. This demonstrates that, depending on the tree depth $L$ and the bias 
probability $p$, either the first two FPTFMs from the same starting point or 
the first FPTFMs from two different initial conditions suffice to uniquely 
determine $p$ and $L$. We note that Fig.\,\ref{Fig:9} in the appendix also 
presents the same analysis as Fig.\,\ref{Fig:3}, but restricted to the regime 
$p\,{\geq}\,0.5$, allowing the use of contour levels with narrower spacing.

\begin{figure*}
\centering
\includegraphics[width=0.85\textwidth]{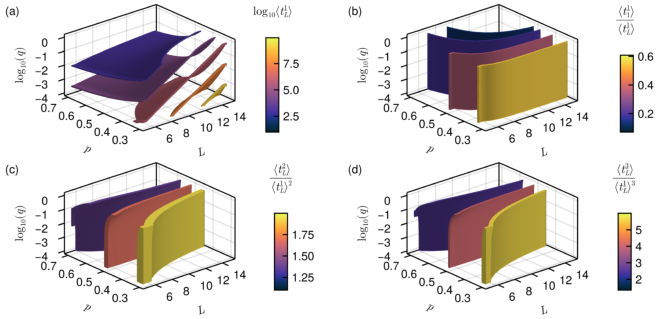}
\caption{(color online) Isosurfaces of constant first passage quantities in the $(L,p,q)$ 
phase space. Each surface corresponds to a fixed value of a given FPT observable, with 
color indicating the specific isovalue. (a) Logarithm of the first factorial moment, 
$\log_{10}\left(\fakm{t_L}{1}\right)$. As predicted by Eq.\,(\ref{Eq:FM_Waiting_first3}), 
all isosurfaces are parallel, reflecting the linear dependence on the first factorial 
moment of the waiting time $\fakm{\tau_w}{1} \,{=}\, \frac 1 q$. (b) Ratio $\frac{
\fakm{t_1}{1}}{\fakm{t_L}{1}}$, comparing first factorial moments for walkers starting 
near the root versus from the leaves. (c) Normalized second factorial moment, $\frac{
\fakm{t_L}{2}}{\fakm{t_L}{1}^2}$. (d) Normalized third factorial moment, $\frac{
\fakm{t_L}{3}}{\fakm{t_L}{1}^3}$. Panels (c) and (d) show that the surfaces quickly 
lose their dependence on the waiting probability $q$ at low values of $q$, making 
them less informative in that regime. Additionally, the similarity in behavior 
suggests that factorial moments beyond the second contribute little new information 
for structural inference.}
\label{Fig:4}
\end{figure*}

A transition in the behavior of the factorial moments occurs at the bias value $p\,
{=}\,0.5$ for deep trees. For example, the first factorial moment starting from the 
deepest level (corresponding to the mean escape time from regular trees of depth $L$ 
studied in \cite{Skarpalezos13}) can be written as
\begin{equation}
\fakm{\mathfrak{t}_L}{1}
= \frac{L}{2p -1} +\frac{1 -p}{(2p -1)^2}\left[\left(\frac{1-p}{p}\right)^L\!\!\!-1\right].
\end{equation}
For deep trees, this expression exhibits two distinct regimes depending on the bias parameter 
$p$: linear scaling with $L$ when $p\,{>}\,\frac{1}{2}$, and exponential growth with $L$ when 
$p\,{<}\,\frac{1}{2}$:
\begin{align}
  \fakm{\mathfrak{t}_L}{1}
  &\stackrel{L \gg 1}{\simeq} \begin{cases}
    \frac{L}{2p -1} &,\, p > \frac 1 2 \\
    \frac{1 -p}{(2p -1)^2} \left(\frac{1-p}{p}\right)^L &,\, p < \frac 1 2
  \end{cases}.
\end{align}
Thus, $p\,{=}\,0.5$ marks the crossover point between qualitatively different scaling behaviors. 
Owing to the recurrence relation (\ref{Eq:FM_recurrence}), it is natural to expect that 
higher factorial moments display a similar crossover at the same bias value.

\section{Random walk with waiting}
\label{Sec:RW_with_waiting}

We now turn to the case in which the random walker experiences waiting times. These 
delays could be caused by traps along the links between nodes, but even in the absence 
of such traps, geometric waiting can serve as a simple way to model diffusive travel 
time between nodes. While we treat the process as discrete hopping, we emphasize that---with 
an appropriate waiting time distribution $w(t)$---all results remain valid if 
the walker moves continuously along the links and the process of the last visited node 
is considered.

A priori, the $k$th FPTFM of a walk with waiting $\fakm{t}{k}$, as given by 
Eq.\,(\ref{Eq:FM_Waiting_first3}), depends on $2k$ quantities: the first $k$ FPTFMs 
of the walk without waiting $\fakm{\mathfrak t}{k}$, and the first $k$ moments of 
the waiting time distribution $\fakm{\tau_w}{k}$. If the waiting time moments are 
known (as also assumed in \cite{Shaebani18}), they can be substituted into 
Eq.\,(\ref{Eq:FM_Waiting_first3}), resulting in a triangular linear system 
of equations. Solving this system for the first two FPTFMs without waiting, 
$\fakm{\mathfrak t}{1}$ and $\fakm{\mathfrak t}{2}$, is straightforward, 
thereby reducing the problem to the previously treated case without waiting.

If the moments of the waiting time distribution are unknown, the number of unknowns 
must be reduced by further assumptions. First, since the FPT distribution of the 
walk without waiting depends only on the two parameters $p$ and $L$, only two of 
its factorial moments $\fakm{\mathfrak t}{k}$ are independent. This limits the 
number of unknowns in Eq.\,(\ref{Eq:FM_Waiting_first3}) to $\min(2, k) \,{+}\, k$. 
To reduce the number of unknowns further---ideally to match the number of equations---one 
must assume or know a specific parametric form of the waiting time distribution 
$w(t)$. If $w(t)$ depends on $\ell$ parameters, then only $\ell$ of its factorial 
moments $\fakm{\tau_w}{k}$ are independent. Therefore, a necessary condition for 
solvability is that $2 \,{+}\, \ell \,{\leq}\, k$. These, however, are necessary 
but not sufficient conditions. As it is demonstrated by the example of a spontaneously 
moving walker in the following subsection, measuring first passage moments beyond 
the second may still be insufficient to uniquely determine all parameters. This 
ambiguity can be resolved by additionally measuring the first moment of the FPT 
for a different initial condition.

\begin{figure*}
\centering
\includegraphics[width=0.7\textwidth]{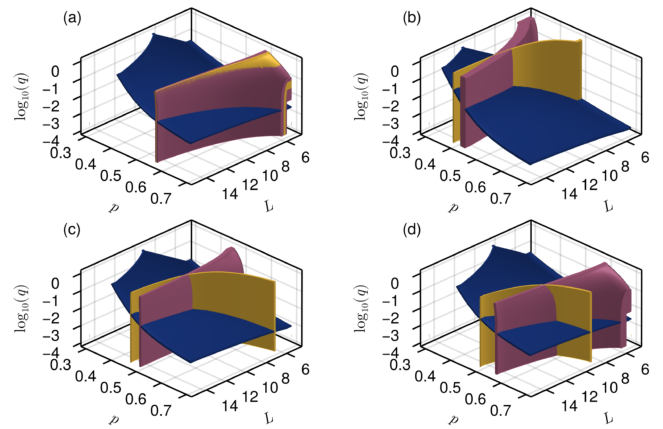}
\caption{(color online) Isosurfaces of constant first passage quantities in the $(L,p,q)$ 
phase space. The intersection point of the three surfaces identifies the parameters $L$, 
$p$, and $q$. This provides a 3D analog to the contour plots shown in Figs.\,\ref{Fig:3}(e) 
and (f). (a) Isosurfaces for the logarithm of the first factorial moment $\log_{10}(
\fakm{t_L}{1})$ (blue), the second normalized factorial moment $\frac{\fakm{t_L}{2}}{
\fakm{t_L}{1}^2}$ (red), and the third normalized factorial moment $\frac{\fakm{t_L}{3}}{
\fakm{t_L}{1}^3}$ (yellow). The values these quantities attain belong to a system with 
$p\,{=}\,0.6$, $L\,{=}\,10$ and $q\,{=}\,0.01$. As the second and third normalized 
moments produce nearly overlapping surfaces, measuring these three quantities does 
not allow for a unique reconstruction of the tree parameters in this regime. (b-d) 
Isosurfaces for $\log_{10}(\fakm{t_L}{1})$ (blue), $\frac{\fakm{t_L}{2}}{\fakm{t_L}{1}^2}$ 
(red), and the ratio $\frac{\fakm{t_1}{1}}{\fakm{t_L}{1}}$ (yellow). The values these 
quantities attain belong to a system with $L\,{=}\,10$, $q\,{=}\,0.01$, and (b) $p\,{=}
\,0.4$, (c) $p\,{=}\,0.5$, and (d) $p\,{=}\,0.6$. In panels (b) to (d), the three surfaces 
intersect at a single point, confirming that the combination of these observables enables 
unambiguous determination of $L$, $p$ and $q$.}
\label{Fig:5}
\end{figure*}

Alternatively, following the approach of \cite{Shaebani18}, instead of measuring 
higher-order FPTFMs, one can match the number of unknowns by measuring lower-order 
factorial moments under several distinct waiting scenarios. Let $n$ be the number 
of different parameter sets considered for the waiting time distribution. Then the 
inequality $\min(2, k) +n\min(\ell, k) \leq n k$ must be satisfied. It is easy to 
see that this inequality can only hold when $k > \ell$, i.e., more moments must 
be measured than there are independent moments of the waiting time distribution. 
Since one can assume that the waiting time distribution depends on at least one 
parameter, this implies that $k\,{\geq}\,2$ and the inequality reduces to $n (k -\ell) 
\,{\geq}\,2$. A detailed exploration of this idea lies beyond the scope of this paper.

\subsection{Example: geometric waiting---Higher moments do not reveal additional information.}
\label{seq:example_spontanous_movement}

In this subsection, the determination of structural parameters of the tree from 
the factorial moments is exemplified using one of the simplest cases: a walker 
that moves spontaneously. That is, at each discrete time step, the walker takes 
a step with probability $q$ or remains at the current node with probability 
$1\,{-}\,q$. The resulting waiting times are geometrically distributed as defined 
in Eq.\,(\ref{Eq:Geometric_Waiting}), and their factorial moments are given by 
$\fakm{\tau_w}{n} \,{=}\, n!\,\frac{(1 -q)^{n-1}}{q^n} \,{=}\, n!\,(1-q)^{n-1}
\fakm{\tau_w}{1}^n$. This system closely resembles the one studied in \cite{Jose18,
Shaebani18}, where the total probability of leaving the deepest level was also 
treated as a tunable parameter.

Figure \ref{Fig:4} shows isosurfaces of the factorial moments in the three-dimensional 
phase space spanned by the structural parameters $p$ and $L$ and the moving probability 
$q$. The isosurfaces of the logarithm of the first FPTFM starting from the deepest 
level $\fakm{t_L}{1}$, shown in Fig.\,\ref{Fig:4}(a), are all parallel, as predicted 
by Eq.\,(\ref{Eq:FM_Waiting_first3}). Furthermore, for a fixed value of $\fakm{t_L}{1}$ 
(e.g., obtained from a measurement), the function $\log_{10}\left(q(p, L)\right)$ is 
proportional to the first FPTFM without waiting, $\fakm{\mathfrak t_L}{1}(p, L)$. 
This suggests that in \cite{Jose18}, the intersection of contour lines of the MFPT 
measured at different laziness levels (i.e., moving probabilities) at a unique 
$(p,L)$ pair is solely due to changes in the probability of leaving the deepest 
level. In that study, these changes were not proportional to changes in the moving 
probability at all other levels. It is also worth noting that $\log_{10}\left(q
\right) \propto -\log_{10}\fakm{\tau_w}{1}$, so a qualitatively similar plot would 
be obtained for any waiting time distribution if $\log_{10}\fakm{\tau_w}{1}$ is 
used as the vertical axis.

In Fig.\,\ref{Fig:4}(b), isosurfaces of the ratio of the first FPTFMs starting 
from the level below the root and from the deepest level are shown. These 
isosurfaces are aligned parallel to the $\log_{10}(q)$-axis, as predicted 
by Eq.\,(\ref{Eq:FM_Waiting_first3}). This behavior arises because, in the 
ratio, the contributions from the waiting time distribution cancel out. As 
a result, the ratio depends solely on the structural parameters of the tree 
and not on the specific form of the waiting time distribution. This cancellation 
property holds generally, regardless of the particular choice of $w(t)$.

From Figs.\,\ref{Fig:4}(c) and (d), which display the second and third normalized 
FPTFMs, it can be observed that these quantities lose their dependence on the 
stepping probability $q$ as $q \,{\rightarrow}\, 0$, i.e., as the mean waiting 
time $\fakm{\tau_w}{1} {\rightarrow}\, \infty$. This behavior can be generalized 
using Eq.\,(\ref{Eq:FM_Waiting_first3}), which shows that it holds for all 
higher normalized factorial moments and that 
\begin{equation}
\frac{\fakm{t}{m}}{\fakm{t}{1}^m} \xrightarrow{q \rightarrow 0} \sum_{\vek b 
\in B_m} \genmultinom{m}{\vek b}\frac{\fakm{\mathfrak t}{\sumnorm{\vek b}}}{
\fakm{\mathfrak t}{1}^m},
\label{Eq:Limit_q0}
\end{equation}
with relative deviation from the limit bounded by $1 -(1 -q)^m$. As $\frac{
\fakm{t}{m}}{\fakm{t}{1}^m}$ converges to a weighted sum of the FPTFMs $\fakm{
\mathfrak t}{k}$ of the walk without waiting, any information about the waiting 
time distribution is effectively lost. This renders normalized factorial moments 
beyond the second ineffective for inferring properties of the waiting dynamics. 
The details of this limiting behavior are provided in the Appendix.

This circumstance is illustrated in Fig.\,\ref{Fig:5}(a), where the isosurfaces 
of $\log_{10}\fakm{t}{1}$, $\frac{\fakm{t_L}{2}}{\fakm{t_L}{1}^2}$ and $\frac{
\fakm{t_L}{3}}{\fakm{t_L}{1}^3}$ are shown for the values they attain in a system 
with $p\,{=}\,0.6$, $L\,{=}\,10$ and $q\,{=}\,0.01$. The isosurfaces of $\frac{
\fakm{t_L}{2}}{\fakm{t_L}{1}^2}$ and $\frac{\fakm{t_L}{3}}{\fakm{t_L}{1}^3}$ 
nearly coincide; thus, if only these quantities are available, the parameters 
of the underlying system cannot be uniquely recovered. However, such measurements 
do constrain the system to a one-dimensional manifold in the $(p, L, q)$ space 
given by the intersection of the $\log_{10}\fakm{t}{1}$-isosurface with the 
$\frac{\fakm{t_L}{2}}{\fakm{t_L}{1}^2}$- or $\frac{\fakm{t_L}{2}}{
\fakm{t_L}{1}^2}$-isosurface.

Since the factorial moments beyond the second fail to provide additional information, 
other quantities---such as the ratio $\frac{\fakm{t_1}{1}}{\fakm{t_L}{1}}$, 
representing the first FPTFMs starting from the highest and deepest levels---must 
be used to uniquely recover the parameters $p$ and $L$ (and $q$). The same limitations discussed 
in Sec.\,\ref{Sec:RW_without_waiting} for the case without waiting also apply here. 
Figures \ref{Fig:5}(b) to (d) show the isosurfaces of $\log_{10}\fakm{t_L}{1}$, 
$\frac{\fakm{t_L}{2}}{\fakm{t_L}{1}^2}$ and $\frac{\fakm{t_1}{1}}{\fakm{t_L}{1}}$, 
each plotted for the values these quantities attain in systems with $L\,{=}\,10$, 
$q\,{=}\,0.01$ and $p\,{=}\,0.4$, $p\,{=}\,0.5$, and $p\,{=}\,0.6$, respectively. 
In all cases, the three isosurfaces intersect at a single point---corresponding 
exactly to the parameter set from which the values were derived. This demonstrates 
that measuring these three quantities suffices to uniquely identify the underlying 
tree parameters.

\begin{figure}
\centering
\includegraphics[width=0.4\textwidth]{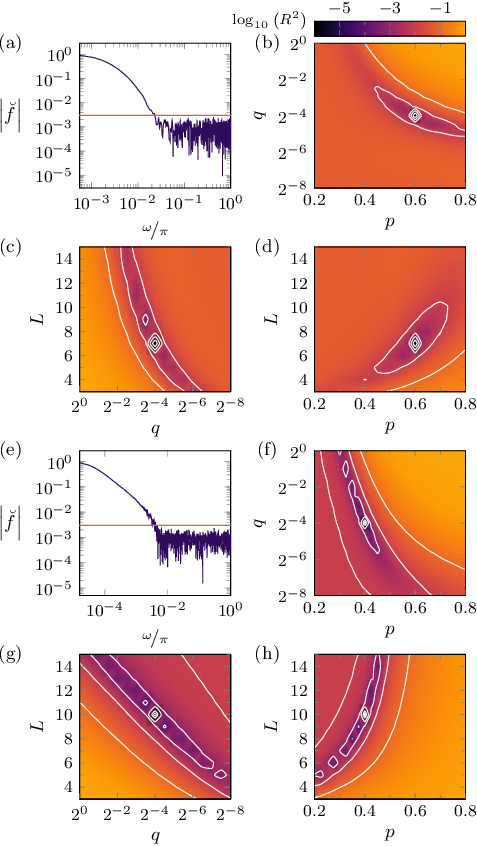}
\caption{(color online) Fourier mode analysis of FPT distributions obtained from Monte Carlo simulations 
for geometrically distributed waiting times. (a) Absolute values of the Fourier modes of 
the FPT distribution, computed from $N\,{=}\,10^6$ trajectories. The horizontal line 
indicates the threshold for selecting significant modes used in the comparison with 
analytical predictions. (b-d) Phase diagrams showing the mean squared deviation $R^2$ 
between the measured and analytical Fourier modes, as a function of model parameters. 
White lines represent contours of constant $R^2$. Default parameters are $p\,{=}\,0.6$, 
$L\,{=}\,7$ and $q\,{=}\,2^{-4}$, unless otherwise varied. (e-h) Same analysis as in 
panels (a-d), but using simulations with $p\,{=}\,0.4$, $L\,{=}\,10$ and $q\,{=}\,2^{-4}$. 
In both cases, $R^2$ exhibits a clear global minimum at the true parameter values used 
for the simulations, indicating the capacity of the method to reliably recover the 
underlying model. Minor local minima are also present but are comparatively shallow.}
\label{Fig:6}
\end{figure}

\begin{figure}
\centering
\includegraphics[width=0.4\textwidth]{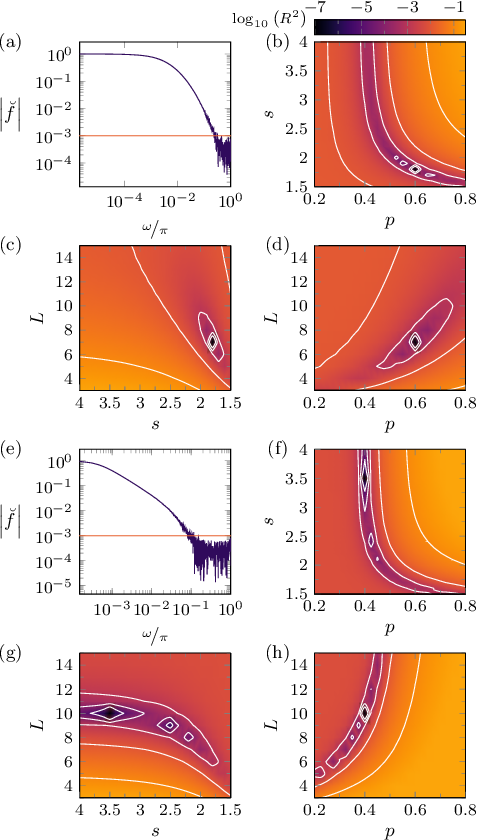}
\caption{(color online) Fourier mode analysis of FPT distributions obtained from Monte Carlo simulations 
for power-law waiting time distributions. (a) Absolute values of the Fourier modes of the 
FPT distribution, computed from $N\,{=}\,10^6$ trajectories. The horizontal line indicates 
the threshold for selecting significant modes used in the comparison with analytical predictions. 
(b-d) Phase diagrams showing the mean squared deviation $R^2$ between the measured and 
analytical Fourier modes, as a function of model parameters. White lines represent contours 
of constant $R^2$. Default parameters are $p\,{=}\,0.6$, $L\,{=}\,7$ and $s\,{=}\,1.8$, 
unless otherwise varied. (e-h) Same analysis as in panels (a-d), but using simulations with 
$p\,{=}\,0.4$, $L\,{=}\,10$ and $s\,{=}\,3.5$. In both cases, $R^2$ exhibits a clear global 
minimum at the true parameter values used for the simulations, indicating the capacity of 
the method to reliably recover the underlying model. Minor local minima are also present 
but are comparatively shallow.}
\label{Fig:7}
\end{figure}

We emphasize that the loss of informativeness of higher moments is not limited to 
cases where the waiting time distribution itself has diverging higher moments. For 
example, in the geometric distribution considered in this section, all factorial 
moments are finite, yet the higher moments of the FPT distribution become 
non-informative when $q$ is small. The $q$-threshold at which this occurs 
increases for larger $L$ or smaller $p$.

\section{Fourier analysis / empirical characteristic function estimation}
\label{Sec:Fourier_method}

As laid out in the previous section, in the presence of waiting, measurements of the 
FPTFMs for a single initial condition are not sufficient to uniquely determine all 
parameters of the random motion. The goal of this section is to overcome this 
shortcoming by employing the Fourier transform of the measured FPT distribution---also 
known as the empirical characteristic function \cite{Cramer46}. This approach 
remains viable even in cases involving fat-tailed waiting time distributions, where 
moments of sufficient order do not exist. In fact, the method of fitting the 
characteristic function has been known for decades and was originally introduced 
to estimate the parameters of stable distributions \cite{Press72,Heathcote77,
Paulson75,Yu04}.

To assess the applicability of this approach, first passage time samples are 
generated using Monte Carlo simulations. The normalized histogram of the simulated 
FPTs is then Fourier transformed using the Fast Fourier Transform (FFT) algorithm, 
yielding estimates of the Fourier transform $\breve f(\omega_k)$ at frequencies 
$\omega_k \,{=}\, \frac{2\pi k}{t_\text{max} +1}$, where $k \,{\in}\, \{0, 1, 
\ldots, t_\text{max}\}$, and $t_\text{max}$ denotes the maximum FPT observed 
in the sample. An analytical expression for the Fourier transform $\tilde 
f_L(\omega)$ is available through the generating function relation $\tilde 
f_L(\omega) \,{=}\, \sum_{t\geq 0} f_L(t) \euler^{\iu\omega t} \,{=}\, \hat 
f_L \left(\euler^{-\iu \omega}\right)$, as provided in Eq.\,(\ref{Eq:GF_Leaves}). 
The objective is to determine the parameters of the random motion by minimizing 
the mean squared deviation
\begin{equation}
R^2 = \frac 1{\abs\Omega} \sum_{\omega \in \Omega} \abs{\tilde f(\omega) 
- \breve f(\omega)}^2,
\end{equation}
where $\Omega$ is the set of frequency modes included in the comparison. This 
metric quantifies the discrepancy between the analytical and empirical Fourier 
transforms.

Since the histogram approximates the FPT distribution with finite accuracy---specifically, 
the number of observed samples at time $t$ follows a Poisson 
distribution with mean $Nf(t)$, where $N$ is the total number of samples---the 
estimated Fourier transform $\breve f$ is subject to statistical fluctuations 
and is not reliable for all frequency modes. A straightforward calculation 
shows that $\Erw{\breve f(\omega)} \,{=}\, \tilde f(\omega)$, and the variance 
satisfies $\Erw{(\breve f(\omega) \,{-}\,\tilde f(\omega))^2} \,{\leq}\, 
\frac 1 {N}$. This implies that indiscriminately including all Fourier modes 
in the error metric $R^2$ is not advisable, as many may be dominated by noise. 
To mitigate this, the frequency set $\Omega$ used in the minimization should 
be restricted to modes with sufficient signal strength:
\begin{equation}
\Omega = \set{\omega_k}{k \in \{0, 1, \ldots, t_\text{max}\} \text{ and } 
\abs{\breve f\left(\omega_k\right)} > \epsilon},
\end{equation} 
where $\epsilon \,{>}\, \frac 1 {\sqrt N}$ is a threshold chosen to exclude 
noisy modes. This approach ensures that the fit prioritizes reliable frequency 
components (see panels (a) and (e) of Figs.\,\ref{Fig:6} and \ref{Fig:7}).

\begin{figure*}
\centering
\includegraphics[width=0.7\textwidth]{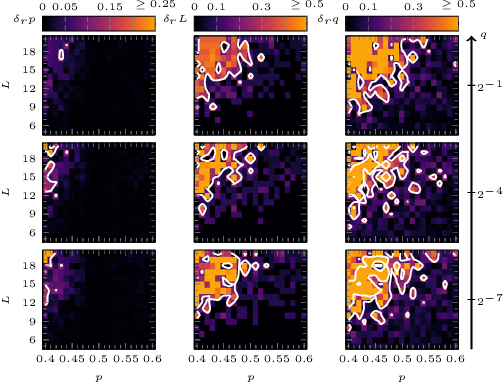}
\caption{(color online) Relative estimation errors for the model parameters $p$, $L$, and $q$ as a function 
of true values of $p$ and $L$, for different levels of moving probability $q$. Each row 
corresponds to a different value of $q$: $q=2^{-1}$ (top), $q=2^{-4}$ (middle), and $q=2^{-7}$ 
(bottom). Columns show the relative deviation for: the upward bias $p$ (first column), the 
tree depth $L$ (second column), and the trap probability $q$ (third column). Contour lines 
indicate levels of relative deviation: For $\delta_r p$, contours are at $0.1$, for $\delta_r L$ 
contours are at $0.2$, and for $\delta_r q$ contours are at $0.2$ and $1.0$. Regions with 
deviations exceeding $0.25$ (for $p$) and $0.5$ (for $L$ and $q$) are shaded in the lightest 
color. Overall, $p$ is reliably estimated across the parameter space. However, estimation 
accuracy for $L$ and $q$ decreases significantly when $p\,{<}\,0.5$, particularly for larger 
tree depths and smaller values of $q$. A strong correlation is observed between the errors 
in $L$ and $q$, suggesting interdependence in their estimation.}
\label{Fig:8}
\end{figure*}

Figures \ref{Fig:6}(b)-(d) and (f)-(g) show the mean quadratic deviation $R^2$ computed 
from numerical data for two parameter sets of a walk with geometrically distributed 
waiting times. Orthogonal slices through the corresponding point in parameter space 
reveal that $R^2$ exhibits a sharp global minimum at the true parameter values. However, 
several local minima also appear, emphasizing the need for careful optimization when 
minimizing $R^2$ to estimate model parameters. This issue becomes more pronounced for 
bias values $p \,{<}\, 0.5$ and larger tree depths $L$. Moreover, regardless of whether 
$p\,{<}\,0.5$ or $p\,{>}\,0.5$, the minimum in the $q$-$L$ section tends to be elongated 
along one direction, indicating that variations in $q$ and $L$ can lead to similar FPT 
distributions.

Figure \ref{Fig:7} presents plots analogous to Fig.\,\ref{Fig:6}, but for a random walk with 
power-law distributed waiting times, $w(t) \,{\propto}\, \frac 1 {t^s}$. Panels (a)-(d) correspond 
to $s\,{=}\,1.8$, where even the mean waiting time does not exist. Panels (e)-(h) show results 
for $s\,{=}\,3.5$, where only the third and higher moments diverge. Despite the heavy tails, 
the mean squared deviation $R^2$ still exhibits a sharp global minimum at the true parameter 
values, though several local minima are also present. In this case, the waiting time exponent 
$s$ and the bias parameter $p$ can compensate for each other, leading to similar FPT distributions 
when varied jointly.

The behavior of the mean quadratic deviation $R^2$ shown in Fig.\,\ref{Fig:6} and Fig.\,\ref{Fig:7} 
is promising, suggesting that the global minimum corresponding to the true parameters of the walk 
can indeed be found using an optimization algorithm. As a proof of concept, $R^2$ was minimized 
for a dataset of $7056$ precomputed histograms of FPTs, covering the parameter ranges $0.4 \,{\leq}\, 
p \,{\leq}\, 0.6$, $5 \,{\leq}\, L \,{\leq}\, 20$ and $2^{-10} \,{\leq}\, q \,{\leq}\, 1$. The 
estimated parameters were then compared to the true parameters used to generate each histogram. 
For this task, differential evolution was employed---an optimization algorithm that does not 
rely on gradient information and is well-suited to finding global minima in the presence of 
local minima \cite{Storn97}. This choice was arbitrary; other optimization strategies may also 
perform well or even better in this context.

Figure \ref{Fig:8} shows the relative deviations of the estimated parameters from 
their true values on $p$-$L$ sections of the phase space for different values of 
the moving probability $q$. The results indicate that the bias probability $p$ 
can be estimated with high accuracy, with relative deviations typically below 
$5\%$. In contrast, accurate estimation of $q$ and $L$ is more sensitive. Low 
deviations in these parameters are generally achieved only when $p\,{>}\,0.5$ 
or when the depth $L$ is small. For $p\,{<}\,0.5$ and sufficiently large $L$, 
the relative deviations in both $L$ and $q$ become significant and tend to 
increase further as $q$ decreases.

We note that the empirical characteristic function approach is indeed a reliable tool 
for parameter estimation in cases where the distribution is only described through its 
characteristic function \cite{Yu04}. Since, for any waiting time distribution with a known 
characteristic function, the characteristic function of the corresponding FPT 
distribution can be derived analytically, we expect the empirical characteristic 
function approach to be in principle generalizable beyond the geometric and power-law 
cases considered here.

\section{Conclusion and outlook}
\label{Sec:Conclusion}

We developed a framework for inferring the structure of nonuniform finite Cayley trees from 
first passage time statistics. We showed that the first two factorial moments of FPTs are 
sufficient to uniquely determine the depth and geometric bias of the tree in the absence 
of trapping effects. When traps or sticky regions are present, we demonstrated that the 
generating function of the FPT distribution is the composition of the generating functions 
of the FPT distribution without waiting and the waiting time distribution. This decomposition 
leads to a nonlinear system of equations connecting factorial moments across different 
scenarios. We further explored strategies to resolve this inverse problem under trapping 
conditions. In particular, we found that varying the initial position of the random walker 
or analyzing the Fourier transform of the FPT distribution allows successful inference 
even for power-law waiting time distributions, where traditional moment-based approaches 
fail due to divergence of higher-order moments.

In the present work, we considered the case of absorbing site being located at the root. 
However, more generally, the target can be placed at an arbitrary node. To discuss the 
possibility of handling such scenarios, we note that previous numerical studies in the 
context of neuronal dendrites \cite{Jose18,Shaebani18} have shown that for moderate 
deviations from a perfectly regular Cayley tree, the MFPT deviates only mildly from 
the effective value obtained for the regular case. This suggests that the inference 
framework should remain reasonably robust even when the absorbing site is not located 
at the root. When the target is placed at another node, two cases for the bias of the 
motion can be distinguished:\ (i) Bias directed towards the target site: In this case, 
the tree can be reordered so that the target becomes the new root. This results in an 
irregular structure, to which the considerations above apply. (ii) Bias directed towards 
the root:\ Here the situation is more involved. Since the level of the particle alone is 
insufficient to determine arrival at the target, the process can no longer be mapped onto 
a simple 1D walk. However, the structure can still be decomposed into a full 
($\mathfrak z\,{-}\,1$)-ary child tree below the target ($\mathfrak z$ being the 
coordination number) and segments between the target and the root, connected to 
($\mathfrak z\,{-}\,1$)-ary subtrees of varying depths. These subtrees can be considered 
as traps, allowing to map the structure between target and the root onto a 1D random 
walk with site-dependent waiting probabilities. Such a decomposition might allow for 
the formulation of renewal-type equations for the FPT distribution to an arbitrary 
target site.

Our analysis revealed that the empirical characteristic function estimation does not always 
provide accurate estimates for the parameters $L$ and $q$ across the full parameter space. 
To mitigate this, we selected a very low threshold value for significant modes, thereby 
retaining more information and reducing the regions where estimation fails. Nevertheless, 
as an interesting avenue for future research, further work is needed to better understand 
which modes are essential for obtaining robust estimates of $p$, $L$ and $q$. 

Our results offer new tools for structural inference across a broad class of branched 
systems---from biological networks such as bronchial, vascular, and neuronal dendritic 
trees to synthetic and engineered systems like communication, utility, and transportation 
networks. The proposed method enables noninvasive characterization of network structures 
in scenarios where direct imaging or tracer tracking is challenging or infeasible. While 
our study focused on tree-like networks, the underlying ideas are extendable to more complex 
architectures, including graphs with loops or hybrid tree-graph topologies. Future extensions 
may incorporate irregular or weighted networks, as well as scenarios involving noisy or 
incomplete first-passage time data. These directions naturally lend themselves to integration 
with statistical inference techniques or machine learning methods, broadening the applicability 
of our framework. By linking random walk dynamics with inverse structural inference, our 
approach adds to the growing repertoire of tools for uncovering hidden features of complex 
systems through stochastic observations. 

We would like to thank Matthias K.\ Hoffmann for fruitful discussions. This work was 
supported by the Deutsche Forschungsgemeinschaft (DFG) within the collaborative research 
center SFB 1027 and also via grants INST 256/539-1, which funded the computing resources 
at Saarland University. R.S.\ acknowledges support by the Young Investigator Grant of 
Saarland University, Grant No.\ 7410110401. \\

\noindent{\bf Appendix: Derivation of Eq.\,(\ref{Eq:Limit_q0}) and the error bound} \\ 

In this Appendix, we derive Eqs.\,(\ref{Eq:FM_Waiting_first3}) and (\ref{Eq:Limit_q0}), as well 
as the estimate for the relative deviation. 

Equation\,(\ref{Eq:FM_Waiting_first3}) expresses the FPTFMs of a random walk with waiting 
in terms of the FPTFMs of the walk without waiting $\fakm{\mathfrak t}{k}$ and the 
factorial moments of the waiting time distribution $\fakm{\tau_w}{k}$. It follows 
from applying Eq.\,(\ref{Eq:FM_S}) to Eq.\,(\ref{Eq:GF_FPT_Waiting}). For the $n$-th 
FPTFM $\fakm{t}{n}$, one requires the $n$-th derivative of a composite function. 
Using the notation of Eq.\,(\ref{Eq:FM_Waiting_first3}), this derivative can be 
written as
\begin{equation}
  \nonumber
  \diff[n]{}{z} \hat{\mathfrak f}(\hat w(z)) =
  \sum_{\vek b \in B_n} M_{\vek b}^n \hat{\mathfrak f}^{(\sumnorm{\vek b})}(\hat w(z))
  \prod_{j=1}^{n}\left({\frac {\hat w^{(j)}(z)}{j!}}\right)^{b_{j}}\!\!\!\!,
\end{equation}
where the superscript in parenthesis denotes higher derivatives, which is a standard form 
of Fa\`a di Bruno's formula, see e.g.\ \cite{Riordan80}. Evaluating this expression at 
$z\,{=}\,1$ gives $\fakm{t}{n}$. Since $\hat w(1) \,{=}\, 1$ by normalization, we have 
$\hat{\mathfrak f}^{(k)}(1) \,{=}\, \fakm{\mathfrak t}{k}$ and $\hat w^{(k)} \,{=}\, 
\fakm{\tau_w}{k}$, which directly yields Eq.\,(\ref{Eq:FM_Waiting_first3}).

Next, we derive Eq.\,(\ref{Eq:Limit_q0}). By applying Eq.\,(\ref{Eq:FM_Waiting_first3}) to 
both the numerator and denominator, and using $\fakm{\tau_w}{1}^m \,{=}\, \prod_{k=1}^m 
\fakm{\tau_w}{1}^{kb_k}$ in the sum, the normalized FPTFM with waiting can be written, 
using the notation introduced in Eq.\,(\ref{Eq:FM_Waiting_first3}), as
\begin{align}
\nonumber \frac{\fakm{t}{m}}{\fakm{t}{1}^m}
&= \sum_{\vek b \in B_m}
  \genmultinom{m}{\vek b}
  \frac{\fakm{\mathfrak{t}}{\sumnorm{\vek b}}}{\fakm{\mathfrak t}{1}^m}
  \frac{\prod_{k=1}^{m}\left(\frac{\fakm{\tau_w}{k}}{k!}\right)^{b_k}}{\fakm{\tau_w}{1}^m} \\ 
&= \sum_{\vek b \in B_m}
  \genmultinom{m}{\vek b}
  \frac{\fakm{\mathfrak{t}}{\sumnorm{\vek b}}}{\fakm{\mathfrak t}{1}^m}
  \prod_{k=1}^{m}\left(\frac{\fakm{\tau_w}{k}}{k!\fakm{\tau_w}{1}^k}\right)^{b_k}\!\!\!\!,
\end{align}
independent of the chosen waiting time distribution.

\begin{figure}[b]
\centering
\includegraphics[width=0.48\textwidth]{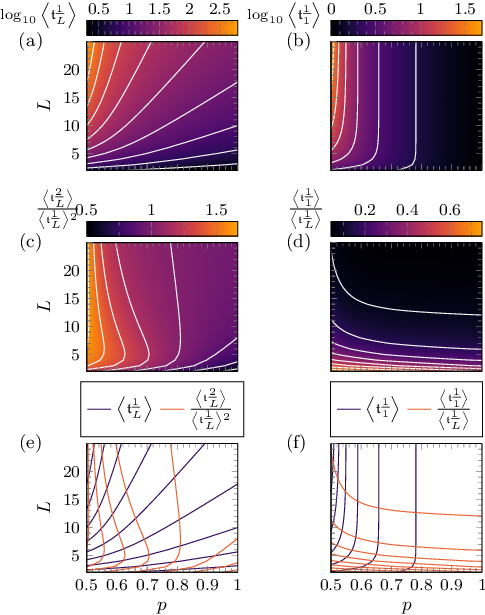}
\caption{(color online) Same analysis as in Fig.\,\ref{Fig:3}, but restricted to the regime  $p\,{\geq}\,0.5$, 
with contour levels adjusted to match the narrower range of deviations.}
\label{Fig:9}
\end{figure}

Recalling that for spontaneous stepping the normalized factorial moments of the waiting 
time distribution are given by $\frac{\fakm{\tau_w}{k}}{\fakm{\tau_w}{1}^k} = k!(1-q)^{k -1}$, 
the expression simplifies further to
\begin{align}
  \nonumber \frac{\fakm{t}{m}}{\fakm{t}{1}^m}
  =& \sum_{\vek b \in B_m}
    \genmultinom{m}{\vek b}
    \frac{\fakm{\mathfrak{t}}{\sumnorm{\vek b}}}{\fakm{\mathfrak t}{1}^m}
    \frac{(1 -q)^m}{(1 -q)^{\sumnorm{\vek b}}} \\
  &\begin{cases}
    \leq \displaystyle\sum_{\vek b \in B_m} \genmultinom{m}{\vek b} 
    \frac{\fakm{\mathfrak{t}}{\sumnorm{\vek b}}}{\fakm{\mathfrak t}{1}^m}, \\
    \geq (1 -q)^m\displaystyle\sum_{\vek b \in B_m} \genmultinom{m}{\vek b} 
    \frac{\fakm{\mathfrak{t}}{\sumnorm{\vek b}}}{\fakm{\mathfrak t}{1}^m}. \\
  \end{cases}
  \label{Eq:Bounds_Norm_FM}
\end{align}
For the ``$\leq$" part of Eq.\,(\ref{Eq:Bounds_Norm_FM}), we use that $\norm{\vek b}_1 \,{\leq}\, 
m$, which implies that every summand is positive and bounded above by $\genmultinom{m}{\vek b} 
\frac{\fakm{\mathfrak{t}}{\sumnorm{\vek b}}}{\fakm{\mathfrak t}{1}^m}$. For the ``$\geq$" part, 
observe that $0 \,{<}\, 1\,{-}\,q \,{\leq}\, 1$, so for any exponent $\nu \,{\in}\, \mathbf N$, 
we have $1 \leq \frac 1 {(1-q)^\nu} < \infty$. Therefore, omitting the factor $\frac{1}{(1 -q)^{
\sumnorm{\vek b}}}$ from each summand decreases its value or leaves it unchanged, which justifies 
the lower bound. From Eq.\,(\ref{Eq:Bounds_Norm_FM}) it follows that
\[ \frac{\fakm{t}{m}}{\fakm{t}{1}^m} \xrightarrow{q \rightarrow 0}\displaystyle\sum_{\vek b \in B_m} 
\genmultinom{m}{\vek b} \frac{\fakm{\mathfrak{t}}{\sumnorm{\vek b}}}{\fakm{\mathfrak t}{1}^m}, \]
and the relative deviation from the limit is given by $1 \,{-}\,(1 \,{-}\,q)^m \stackrel{q \ll 1}{\sim} mq$.

As $\fakm{\mathfrak t}{1} {\rightarrow} \infty$, the normalized factorial moments $\frac{\fakm{t}{m}}{
\fakm{t}{1}^m}$ remain bounded (c.f.\ Fig.\,\ref{Fig:3} and Fig.\,\ref{Fig:4}(c),(d)), implying that 
$\fakm{\mathfrak t}{m} \sim \fakm{\mathfrak t}{1}^m$ in this limit. Since $\sumnorm{\vek b} \leq m$ 
for all $\vek b \in B_m$, it follows that
\[ \lim_{q\rightarrow 0}\frac{\fakm{t}{m}}{\fakm{t}{1}^m} \stackrel{\fakm{\mathfrak t}{1} \gg 1}{\sim} 
\frac{\fakm{\mathfrak t}{m}}{\fakm{\mathfrak t}{1}^m} \]
in the small $q$ limit. This explains why in Figs.\,\ref{Fig:4}(b),(d) the isosurfaces appear nearly 
invariant with respect to $q$ when $p\,{<}\,0.5$.

\bibliography{Refs-CayleyTree} 

\end{document}